\documentclass[aip,jcp,preprint,groupedaddress]{revtex4-1}
 
\usepackage{graphicx}
\usepackage{dcolumn}
\usepackage{bm}
\usepackage{amsmath}
\usepackage{textcomp}
\usepackage{longtable}
\usepackage{url}
\usepackage{tabularx}
\usepackage{threeparttable}
\usepackage{multirow}

\usepackage[bookmarks=false, hidelinks, linktocpage=true]{hyperref}

\begin{document}

\title{Elastic properties of LaNiO$_3$ from first-principles calculations}

\author{\v{S}. Masys }
\email[Corresponding author. Email address: ]{Sarunas.Masys@tfai.vu.lt}

\author{V. Jonauskas}

\affiliation{Institute of Theoretical Physics and Astronomy, Vilnius University, A. Go\v{s}tauto Street 12, LT-01108 Vilnius, Lithuania}

\date{\today}

\begin{abstract}
By applying density functional theory (DFT) approximations, we present a first-principles investigation of elastic properties for the experimentally verified phases of a metallic perovskite LaNiO$_3$. In order to improve the accuracy of calculations, at first we select the most appropriate DFT approaches according to their performance in reproducing the low-temperature crystalline structure and the electronic density of states observed for the bulk LaNiO$_3$. Then, we continue with the single-crystal elastic constants and mechanical stability for the most common rhombohedral as well as high-temperature cubic and strain-induced monoclinic phases. Together with the calculated single-crystal elastic constants, the deduced polycrystalline properties, including bulk, shear, and Young's moduli, Poisson's ratio, Vickers hardness, sound velocities, Debye temperature, and anisotropy indexes, remedy the existing gap of knowledge about the elastic and mechanical behaviour of LaNiO$_3$, at least from a theoretical standpoint.           
 
\end{abstract}


\maketitle

\section{Introduction}

Among the remarkable family of perovskite oxides, lanthanum nickelate LaNiO$_3$ is a rare example characterized by paramagnetic metallic behaviour down to the lowest temperatures and being structurally compatible with many active functional layers [\onlinecite{dobin_1}, \onlinecite{gou_2}]. For a long time, the technological interest in this material has been limited to the development of highly conductive bottom electrodes for various ferroelectric thin-film devices [\onlinecite{murugavel_3, jan_4, kim_5}]. But recently, the enthusiasm was renewed by experimental recognition of electric-field tunable metal-to-insulator transition and theoretical findings that heterostructures and superlattices composed of thin LaNiO$_{3}$ layers separated by wide-gap insulators could possibly exhibit high-temperature superconductivity [\onlinecite{zubko_6, hansmann_7, chaloupka_8}]. Of course, the fact that the hunt for novel superconducting materials is at the vanguard of condensed matter physics research sheds a new light on the outlook of LaNiO$_{3}$, but on the other hand, it does not automatically mean that the bulk properties of this oxide are already fully explored. For example, although one can find a variety of theoretical studies [\onlinecite{nohara_9, guan_10, masys_11, gou_2, he_13}] based on conventional density functional theory (DFT) and beyond-DFT techniques in which the electronic structure of LaNiO$_{3}$ is extensively investigated and compared to the experimental data, in the literature scarcely any first-principles calculations on the elastic properties of LaNiO$_{3}$ have been reported so far. The situation is not much better with the experiment, especially regarding the single-crystal measurements. Having in mind that the list of potential applications of LaNiO$_{3}$ continues to grow [\onlinecite{zhang_14, hsiao_15, luo_16}], it becomes obvious that the deep knowledge of elastic features could be useful for both technological and fundamental reasons.  

The main goal of our work is to evaluate single-crystal elastic constants, polycrystalline moduli, and other mechanical properties of LaNiO$_{3}$ using first-principles calculations. In order to select the most appropriate DFT-based approaches for it, at first we focus attention on the geometrical parameters and electronic structure of rhombohedrally distorted phase (space group $R\bar{3}c$, No. 167) which is observed in a wide range of temperatures. Then, we continue with the elastic parameters and extend our study to the high-temperature cubic phase ($Pm\bar{3}m$, No. 221) [\onlinecite{zhou_17}] together with monoclinic ($C2/c$, No. 15) [\onlinecite{may_18}] structure that is available under tensile or compressive strains induced by various substrates. We hope that the obtained data will be helpful for many practical applications related to the mechanical behaviour of bulk LaNiO$_{3}$, thin films, and heterostructures.  

\section{Computational details}

All the calculations reported in this paper were performed with the CRYSTAL14 program [\onlinecite{crystal14_24}] which expands the crystalline orbitals as linear combinations of atom-centred Gaussian-type functions. A small-core Hay-Wadt pseudopotential [\onlinecite{hay_26}] was used for describing the core electrons ($1s^{2}2s^{2}2p^{6}3s^{2}3p^{6}3d^{10}4s^{2}4p^{6}4d^{10}$) of La atom, while the valence part ($5s^{2}5p^{6}5d^{1}6s^{2}$) of this basis set was adopted from LaMnO$_{3}$ study [\onlinecite{La_baze_25}]. Concerning the description of Ni and O atoms, the all-electron triple-zeta plus polarization quality basis sets, specifically derived for periodic solid-state treatment, were taken from [\onlinecite{tzvp_27}].

For the sake of simplicity, we have modelled the paramagnetic behaviour of LaNiO$_{3}$ by non-magnetic calculations. The default values of parameters that define the convergence threshold on total energy and truncation criteria for bielectronic integrals were set to more severe ones: (8) and (8 8 8 8 16), respectively. In addition, while performing full geometry optimization the allowed root-mean-square values of energy gradients and nuclear displacements were correspondingly tightened to 0.00006 and 0.00012 a.u. In order to improve the self-consistence field convergence, the Kohn-Sham matrix mixing technique (at 80$\%$) together with Anderson's method [\onlinecite{anderson_28}] were applied. The shrinking factor of 32, 16, and 8 was used for reciprocal space sampling, resulting in 969, 417, and 150 independent \textbf{\textit{k}} points in the first irreducible Brillouin zone for cubic, rhombohedral, and monoclinic phases of LaNiO$_{3}$, respectively. All the other important settings for calculations were left at their default values, which in turn can be found in CRYSTAL14 user's manual [\onlinecite{crystal14_39}].  

A variety of DFT approaches was tested in order to find out which of them exhibit the closest resemblance to the low-temperature experimental data on crystalline structure: starting from standard local density (LDA [\onlinecite{lda_29}, \onlinecite{lda_30}]) and generalized gradient (PBE [\onlinecite{pbe_31}]) approximations up to revised functionals for solids (PBEsol [\onlinecite{pbesol_32}], SOGGA [\onlinecite{sogga_33}], and WC [\onlinecite{wc_34}]) and hybrids (mB1WC, PBE0 [\onlinecite{pbe0_36}], and HSE06 [\onlinecite{hse06_37}]). Note that mB1WC stands for modified B1WC approach [\onlinecite{b1wc_35}] since instead of PW functional [\onlinecite{pw_38}] we have employed the correlation part from PBE framework. The revised functionals, SOGGA and WC, were also combined with PBE correlation. 

A fully automated procedure ELASTCON [\onlinecite{elastcon_40}, \onlinecite{elastcon_41}] developed for calculating single-crystal elastic constants $C_{ij}$ has been applied for all three phases of LaNiO$_{3}$. For each independent strain, two deformed structures with the magnitude of the strain reaching -0.01 and 0.01 were considered. The internal atomic positions were allowed to relax with the deformed cell shape and volume remaining fixed. Polycrystalline elastic properties, starting from the bulk and shear moduli, were derived from the computed single-crystal elastic constants $C_{ij}$ using the Voigt [\onlinecite{voigt_42}] and Reuss [\onlinecite{reuss_43}] approximations. The mean values of these parameters calculated via the averaging method proposed by Hill [\onlinecite{hill_44}] have been utilized to evaluate Young's modulus, Poisson's ratio, sound velocities, and Debye temperature according to well-known expressions that can be found elsewhere [\onlinecite{grimvall_45}, \onlinecite{connetable_47}]. For comparison purposes, the bulk modulus, denoted as $B_{\text{BM}}$, has been obtained by fitting the total energy as a function of volume to the third-order Birch-Murnaghan (BM) equation of state [\onlinecite{birch_46}]. The fitting was done at 11 points within the range of 0.92-1.08 variation of the initial volume by another automated procedure, named EOS, which is also implemented in the CRYSTAL14 code.       

\section{Results and discussion}

\subsection{Performance of DFT approaches}

\begin{figure}
\centering
{
\includegraphics[scale=0.40]{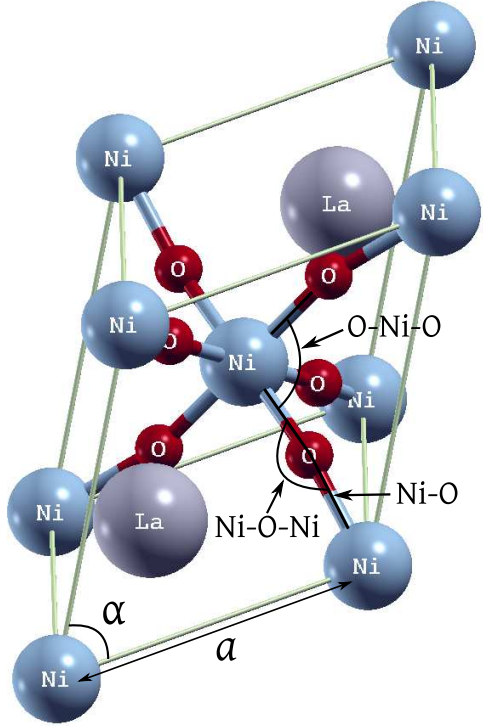}
}
\caption{\label{fig1}
The crystalline structure of rhombohedral ($R$\={3}$c$) LaNiO$_{3}$. The structural parameters are presented in Table \ref{tab1}.}
\end{figure}

Since many specific material properties like elastic constants, sound velocities, and Debye temperature are critically dependent on change in geometry of the system, we primarily focus our attention on the accuracy of the calculated lattice constants and other structural parameters that are shown in Fig. \ref{fig1}. The obtained values together with low-temperature experimental data for rhombohedral LaNiO$_3$ are presented in Table \ref{tab1}. The mean absolute relative error (MARE) was evaluated according to the expression
\begin{equation}
\text{MARE} = \frac{100}{n}\displaystyle\sum\limits_{i=1}^{n}{\bigg \vert \frac{p_{i}^{\text{Calc.}}-p_{i}^{\text{Expt.}}}{p_{i}^{\text{Expt.}}} \bigg \vert},
\label{eq:equ1}
\end{equation}
in which $p_{i}^{\text{Calc.}}$ and $p_{i}^{\text{Expt.}}$ are the calculated and experimental values of the considered parameter, respectively. Note that we did not apply the zero-point anharmonic expansion (ZPAE) corrections for the experimental data. On one hand, the ZPAE corrections are straightforwardly applicable only for the cubic systems [\onlinecite{alchagirov_21}]. On the other hand, ZPAE can expand the equilibrium lattice constant by $1\%$ for light atoms like Li and much less for heavy atoms [\onlinecite{hao_22}], thus it should not have a noticeable influence on material like LaNiO$_{3}$. What is more, we did not analyse the geometry of cubic and monoclinic phases of LaNiO$_{3}$ because of the thermal expansion of the former and internal strains of the latter meaning that the direct comparison between experiment and fully-relaxed zero-temperature calculations would not be accurate.

As it could be expected, Table \ref{tab1} reveals that LDA and PBE functionals tend to underestimate and overestimate lattice constant $a$ together with angle $\alpha$, respectively. As a direct consequence, this tendency is also reflected in a deviation of the volume $V$ values. However, the following results are in a good agreement with the recent plane-wave calculations [\onlinecite{gou_2}, \onlinecite{he_13}] demonstrating the reliability of our chosen basis set. Although the revised functionals -- PBEsol, SOGGA, and WC -- appear worse than PBE for the evaluation of the Ni-O-Ni angle, their overall performance is systematically improved, especially for the aforementioned structural parameters, $a$, $\alpha$, and $V$. The decrease of MARE from 1.15$\%$ for PBE to 0.74$\%$ for SOGGA and impressive numbers of 0.41 and 0.35$\%$ for PBEsol and WC, correspondingly, leaves no doubt about the effectiveness of the modifications made to the revised density functionals for solids. A slightly worse result of SOGGA could be influenced by its different analytical form or more tightly bounded exchange [\onlinecite{sogga_33}]. 

\begin{table}
\caption{\label{tab1}
Calculated structural parameters of rhombohedral LaNiO$_{3}$ compared to the experimental data [\onlinecite{garcia_23}] at 1.5 K. Lattice constant $a$ and bond distance Ni-O are given in \AA, volume $V$ is given in \AA$^{3}$, angles $\alpha$, Ni-O-Ni, and O-Ni-O are given in degrees. MARE (in $\%$) stands for the mean absolute relative error. The numbers in brackets (in $\%$) represent absolute relative errors for each of the considered parameters.      }
\footnotesize
\begin{tabularx}{\textwidth}{>{\hsize=1.4\hsize}X>{\centering\arraybackslash\hsize=0.96\hsize}X>{\centering\arraybackslash\hsize=0.96\hsize}X>{\centering\arraybackslash\hsize=0.96\hsize}X>{\centering\arraybackslash\hsize=0.96\hsize}X
>{\centering\arraybackslash\hsize=0.96\hsize}X>{\centering\arraybackslash\hsize=0.96\hsize}X>{\centering\arraybackslash\hsize=0.96\hsize}X>{\centering\arraybackslash\hsize=0.96\hsize}X>{\centering\arraybackslash\hsize=0.96\hsize}X}
 \hline \hline  
  &  LDA   &    PBE &  PBEsol  &  SOGGA   &    WC &  mB1WC  & PBE0   &    HSE06   & Expt.\\ 
\hline
$a$  &   5.332  &   5.444  &   5.375 &    5.361     &  5.379       &   5.363 &    5.396     &   5.399           &   5.384  \\
   &   (0.95) &   (1.11) &   (0.16)&    (0.42)     &  (0.09)       & (0.39)  &   (0.23)      &     (0.27)         &          \\
$\alpha$  &   60.68  &   61.09  &   60.89 &   60.83   &    60.90    &   60.42 &  60.44  &  60.49  &     60.86 \\
         &   (0.30) &   (0.37) &  (0.05) &    (0.05)     &   (0.06)      &   (0.73)      &  (0.69)             &   (0.60)      &      \\
$V$  &   108.86  &   116.85  &   112.03 &  110.99  &  112.25  &   110.07 &  112.20  &  112.49  &    112.48 \\
   &    (3.22) &   (3.89)  &   (0.40) &  (1.32)    &  (0.20)    & (2.14)      &  (0.25)               &   (0.01)       &          \\
Ni-O  &   1.904     &   1.958  &   1.926 &  1.919  &  1.927  &   1.908 &  1.922  &  1.924  &    1.933 \\
      &   (1.50)       &   (1.29) &  (0.37) &  (0.73)  &  (0.31)   & (1.27)  &   (0.58)             &    (0.45)     &         \\
Ni-O-Ni  &  169.05     &   164.77  &   167.05 &  167.60  & 167.04  &   170.70 &   169.92  & 169.38    &   164.82 \\
         &    (2.57)       &   (0.03)  &  (1.35) &  (1.69)  & (1.35)  & (3.57)  &    (3.09)             &    (2.77)      &          \\
O-Ni-O  &   89.16     &   88.58  &   88.88 & 88.96   &  88.88  &   89.46 &  89.40  &  89.33    &   88.78 \\
        &    (0.43)   &   (0.22) &  (0.12)  &  (0.21)   & (0.12)    & (0.77)  &   (0.71)             &   (0.63)      &         \\
MARE    &    1.49      &   1.15   &  0.41  &  0.74   &   0.35  &   1.48      &   0.92            &   0.79      &         \\    

\hline \hline
\end{tabularx}
\end{table}  

One can note that the inclusion of the standard portion (25$\%$) of the exact Hartree-Fock (HF) exchange -- as it is in the PBE0 scheme -- has a positive impact on the overall performance of PBE functional, since MARE now gets reduced to 0.92$\%$. Furthermore, the rejection of the long-range HF-type exchange in the HSE06 framework seems to have even a better effect. As HSE06 outperforms PBE0 for nearly every structural parameter of rhombohedrally distorted LaNiO$_3$, its MARE takes a value of 0.79$\%$ which is very close to the result of SOGGA. Interestingly, HSE06 functional manages to almost ideally reproduce the experimentally determined value of volume $V$, although neither lattice constant $a$ nor angle $\alpha$ alone are improved in comparison to what was obtained using PBEsol or WC. In this case, HSE06 benefits from an error cancellation which, according to the expression  
\begin{equation}
V=a^{3}\sqrt{1+\cos^{2}\alpha(2\cos\alpha-3)},
\label{eq:equ2}
\end{equation}
is likely if $a$ and $\alpha$ are correspondingly overestimated and underestimated (or vice versa) in respective proportions. 

Another popular hybrid functional for the solid-state calculations, named B1WC, incorporates 16$\%$ of the exact full-range HF exchange. In this work, a modified version mB1WC with PBE correlation was applied in order to find out the effect of HF exchange on solely the WC framework. Unfortunately, Table \ref{tab1} apparently indicates that mB1WC not only degrades the accuracy of its predecessor WC for all the structural parameters, but also is the worst performer among the considered hybrids with MARE reaching 1.48$\%$. It becomes obvious that the inclusion of some part of HF exchange is harmful for WC and, most probably, PBEsol and SOGGA functionals which already demonstrate reliable results in reproducing the crystalline structure of rhombohedral LaNiO$_3$. 

On the whole, the evaluated mean discrepancies between different DFT calculations and experiment do not exceed 1.5$\%$, and in most cases this limit may seem to be acceptable. However, for the sake of greater accuracy, we tighten functional selection criterion to 1$\%$, therefore LDA, PBE, and mB1WC appear to be the ones that should be eliminated before making further calculations of elastic properties.

\begin{figure}
\centering
{
\includegraphics[scale=0.45]{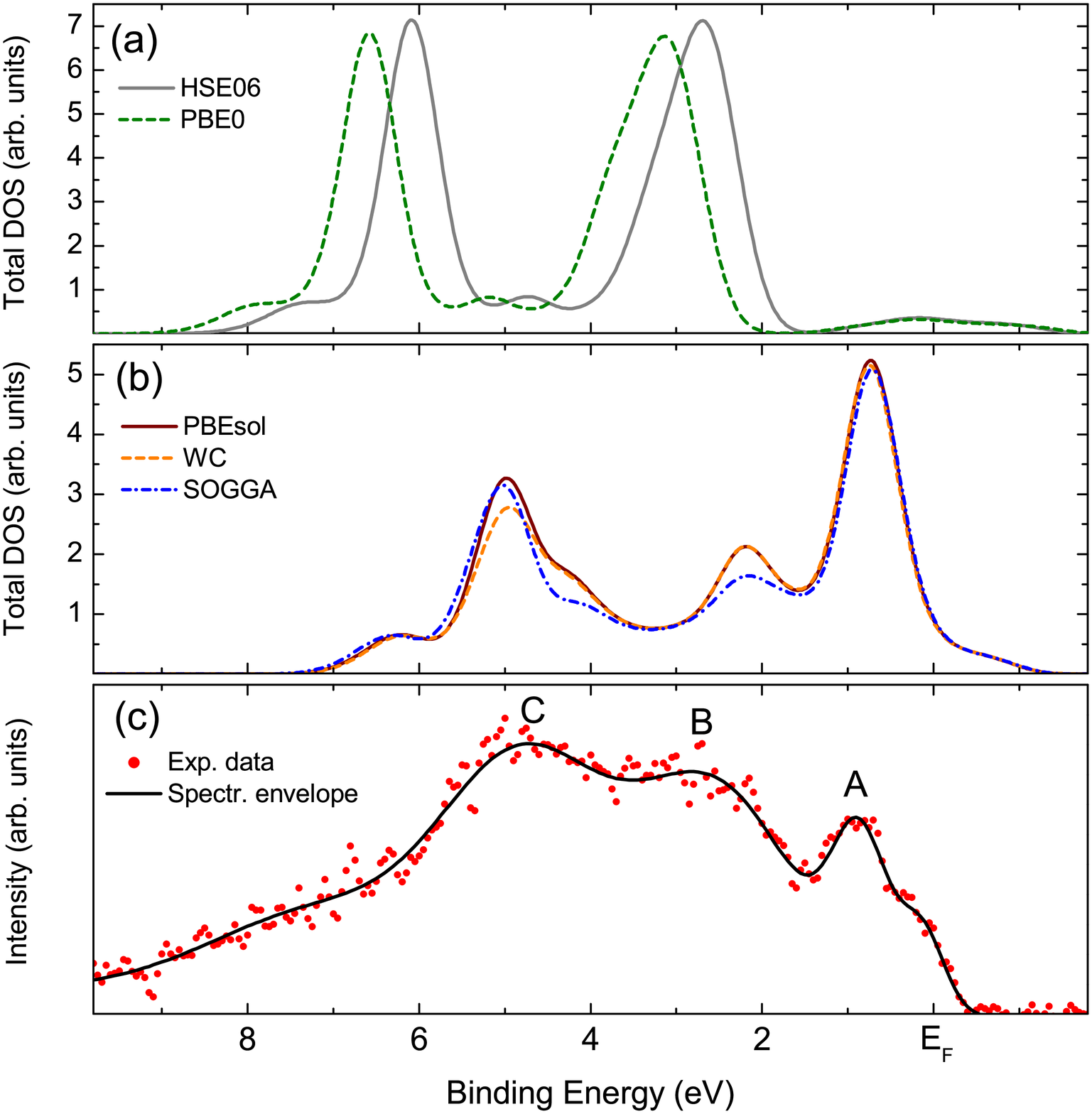}
}
\caption{\label{fig2}
The comparison of total DOS calculated using (a) hybrid and (b) pure DFT approaches with (c) X-ray photoelectron spectroscopy LaNiO$_{3}$ spectrum [\onlinecite{masys_11}]. The Fermi energy is set at zero. }
\end{figure} 

But this is not the end of the story. Since simply reproducing the correct ground-state crystalline structure is not a sufficient criterion to properly evaluate the quality of the DFT approach performance [\onlinecite{gou_2}], in Fig. \ref{fig2} we also compare the density of states (DOS) obtained using the remaining functionals with experimentally observed valence band of LaNiO$_3$. Here, it is clearly seen that hybrids fail to correctly describe the electronic structure exhibiting only two significantly shifted spectral features, while revised functionals demonstrate a fairly good agreement with the main energy peak positions A, B, and C. Needless to say, such combination of accurately reproduced electronic and crystalline structures shown by PBEsol, SOGGA, and WC automatically captures our sympathy. Besides, if one would take into account the computational time and resources required for the calculations, they would have another advantage over the hybrid scheme of PBE0 and HSE06. Thus, to sum up, it can be stated that the revised functionals for solids appear to be the best choice among the considered DFT approximations for the next calculations of elastic properties of LaNiO$_3$. And we hope that the quality of the description of the other two phases will remain at the same level as it is now for rhombohedral LaNiO$_{3}$.    

\subsection{Elastic properties}   

\begin{table}
\caption{\label{tab2}
Single-crystal elastic constants $C_{ij}$ (in GPa) calculated for different phases of LaNiO$_3$.     }
\footnotesize
\begin{tabularx}{\textwidth}{>{\hsize=1.1\hsize}X>{\centering\arraybackslash\hsize=1.1\hsize}X>{\centering\arraybackslash\hsize=1.1\hsize}X>{\centering\arraybackslash\hsize=1.1\hsize}X>{\centering\arraybackslash\hsize=0.5\hsize}X
>{\centering\arraybackslash\hsize=1.1\hsize}X>{\centering\arraybackslash\hsize=1.1\hsize}X>{\centering\arraybackslash\hsize=1.1\hsize}X>{\centering\arraybackslash\hsize=0.5\hsize}X>{\centering\arraybackslash\hsize=1.1\hsize}X>{\centering\arraybackslash\hsize=1.1\hsize}X>{\centering\arraybackslash\hsize=1.1\hsize}X}
 \hline \hline  
  &  \multicolumn{3}{c}{Cubic ($Pm\bar{3}m$)}   &  &  \multicolumn{3}{c}{Rhombohedral ($R\bar{3}c$)} & & \multicolumn{3}{c}{Monoclinic ($C2/c$)}  \\ 
\cline{2-4} \cline{6-8} \cline{10-12}
                    & PBEsol & SOGGA   &    WC  &  &  PBEsol &  SOGGA &   WC    & & PBEsol &SOGGA &  WC  \\
\hline
$C_{11}$            & 405.5  &  415.2  &  405.3 &  &  264.6  &  268.6 &   266.2 & &  341.9 & 348.1 & 342.9 \\
$C_{12}$            & 119.2  &  120.0  &  120.7 &  &  161.6  &  165.1 &   162.4 & &  156.3 & 164.6 & 159.3 \\
$C_{13}$            &        &         &        &  &  150.1  &  151.3 &   151.2 & &  136.5 & 139.6 & 137.3 \\
$C_{15}$            &        &         &        &  &  17.6   &  17.6  &   16.9  & &  -3.9  & -3.7  & -3.8  \\
$C_{22}$            &        &         &        &  &         &        &         & &  269.3 & 270.4 & 270.7 \\
$C_{23}$            &        &         &        &  &         &        &         & &  145.1 & 148.3 & 148.6 \\
$C_{25}$            &        &         &        &  &         &        &         & &  16.7  & 17.2  & 16.9  \\
$C_{33}$            &        &         &        &  &  341.1  &  348.9 &   342.2 & &  301.1 & 307.2 & 302.8 \\
$C_{35}$            &        &         &        &  &         &        &         & &  -30.2 & -30.9 & -29.9 \\
$C_{44}$            & 103.6  &  105.5  &  103.5 &  &  98.8   &  101.6 &   99.5  & &  46.2  & 45.9  & 46.9  \\
$C_{46}$            &        &         &        &  &         &        &         & &  0.7   & -1.7  & -1.2  \\
$C_{55}$            &        &         &        &  &         &        &         & &  82.1  & 83.9  & 82.8  \\ 
$C_{66}$            &        &         &        &  &  51.5   &  51.7  &   51.9  & &  104.9 & 107.8 & 105.7 \\ 
 
\hline \hline
\end{tabularx}
\end{table}   

Single-crystal elastic constants $C_{ij}$ of LaNiO$_3$ calculated by applying revised functionals are presented in Table \ref{tab2}. Unfortunately, no experimental or theoretical data are currently available to deal with, therefore we can only make a comparison between our own results. From Table \ref{tab2} it is seen that PBEsol, SOGGA, and WC perform quite similarly, with SOGGA on average giving slightly higher values than PBEsol and WC. This indicates that the obtained results are reliable and could be useful for various practical applications. Concerning the elastic properties of LaNiO$_3$, one can note that rhombohedral phase is more resistive against uniaxial deformation along the $z$ direction, while monoclinic phase is harder to deform along the $x$ axis. The average value of all three functionals for $C^{\text{Rhomb.}}_{33}$ and $C^{\text{Mon.}}_{11}$ reaches $\sim344$ GPa. The resistance against uniaxial compression or tension along the $y$ direction is found to be practically the same for both structures. Although cubic phase exhibits the highest incompressibility of all under the uniaxial strains with $C^{\text{Cub.}}_{11}=C^{\text{Cub.}}_{22}=C^{\text{Cub.}}_{33}\approx409$ GPa, its resistance against shear deformations is reduced to $C^{\text{Cub.}}_{44}=C^{\text{Cub.}}_{55}=C^{\text{Cub.}}_{66}\approx104$ GPa. A similar value is demonstrated by rhombohedral structure for shear deformations around the $x$ and $y$ axes ($C^{\text{Rhomb.}}_{44}=C^{\text{Rhomb.}}_{55}\approx100$ GPa) and by monoclinic phase for shear deformation around the $z$ axis ($C^{\text{Mon.}}_{66}\approx106$ GPa). But despite that, the latter is much easier to deform around the $x$ axis with $C^{\text{Mon.}}_{44}\approx46$ GPa, whereas the former -- around the $z$ axis with $C^{\text{Rhomb.}}_{66}\approx52$ GPa. Some of the elastic constants, such as $C^{\text{Mon.}}_{15}$ or $C^{\text{Mon.}}_{35}$, are negative and this may imply that monoclinic phase of LaNiO$_3$ could be mechanically unstable. However, this is not the case, since the stability restrictions on $C_{ij}$, following from the requirement that the strain energy must always be positive for all applied strains, are fully satisfied. It indicates that cubic, rhombohedral, and monoclinic structures of LaNiO$_3$ are mechanically stable at zero temperature and zero pressure conditions. The appropriate stability criteria on $C_{ij}$ for different phases of material can be found elsewhere [\onlinecite{wu_48}, \onlinecite{tapia_49}]. 

\begin{table}
\caption{\label{tab3}
Polycrystalline bulk modulus (in GPa) from Voigt ($B_{\text{V}}$), Reuss ($B_{\text{R}}$), and Hill ($B$) approximations, bulk modulus $B_{\text{BM}}$ (in GPa) and its pressure derivative $B'_{\text{BM}}$ from BM equation of state, shear modulus (in GPa) from Voigt ($G_{\text{V}}$), Reuss ($G_{\text{R}}$), and Hill ($G$) approximations, Young's modulus $E$ (in GPa), Poisson's ratio $\nu$, longitudinal sound velocity $v_{\text{L}}$ (in m/s), transverse sound velocity $v_{\text{T}}$ (in m/s), average sound velocity $\bar{v}$ (in m/s), Debye temperature $\theta_{\text{D}}$ (in K), Vickers hardness $H_{\text{V}} $ (in GPa), $B/G$ ratio, anisotropy index in compressibility $A_{B}$ (in $\%$), anisotropy index in shear $A_{G}$ (in $\%$), and universal anisotropy index $A^{\text{U}}$ calculated for different phases of LaNiO$_3$.     }
\footnotesize
\begin{tabularx}{\textwidth}{>{\hsize=1.1\hsize}X>{\centering\arraybackslash\hsize=1.1\hsize}X>{\centering\arraybackslash\hsize=1.1\hsize}X>{\centering\arraybackslash\hsize=1.1\hsize}X>{\centering\arraybackslash\hsize=0.5\hsize}X
>{\centering\arraybackslash\hsize=1.1\hsize}X>{\centering\arraybackslash\hsize=1.1\hsize}X>{\centering\arraybackslash\hsize=1.1\hsize}X>{\centering\arraybackslash\hsize=0.5\hsize}X>{\centering\arraybackslash\hsize=1.1\hsize}X>{\centering\arraybackslash\hsize=1.1\hsize}X>{\centering\arraybackslash\hsize=1.1\hsize}X}
 \hline \hline  
  &  \multicolumn{3}{c}{Cubic ($Pm\bar{3}m$)}   &  &  \multicolumn{3}{c}{Rhombohedral ($R\bar{3}c$)} & & \multicolumn{3}{c}{Monoclinic ($C2/c$)}  \\ 
\cline{2-4} \cline{6-8} \cline{10-12}
                    & PBEsol & SOGGA   &    WC  &  &  PBEsol &  SOGGA &   WC    & & PBEsol &SOGGA &  WC  \\
\hline

$B_{\text{V}} $     &        &         &        &  &  199.3  &  202.4 &  200.5  & &  198.7 & 203.4 & 200.8  \\ 
$B_{\text{R}} $     &        &         &        &  &  197.5  &  200.6 &   198.6 & &  196.7 & 201.2 & 199.0 \\ 
$B$                 & 214.7  &  218.4  &  215.6 &  &  198.4  &  201.5 &   199.6 & &  197.7 & 202.3 & 199.9 \\
$B_{\text{BM}}$    & 214.7  &  218.5  &  215.6 &  &  198.9  &  202.5 &   200.0 & &  199.3 & 201.9 & 199.3 \\
$B'_{\text{BM}}$   &  4.26  &  4.28   &  4.25  &  &  4.28   &  4.26  &   4.28  & &  4.22  & 4.26  & 4.22  \\
$G_{\text{V}} $     & 119.4  & 122.3   &  119.0 &  &  77.1   & 78.8   &  77.5   & &  78.3  & 79.1  & 78.5  \\ 
$G_{\text{R}} $     & 116.5  &  119.0  &  116.2 &  &  66.7   &  68.0  &   67.6  & &  69.1  & 69.1  & 69.2 \\
$G$                 & 118.0  &  120.7  &  117.6 &  &  71.9   &  73.4  &  72.6   & &  73.7  & 74.1  & 73.9  \\ 
$E$                 & 299.1  &  305.7  &  298.5 &  &  192.6  &  196.5 &   194.2 & &  196.6 & 198.0 & 197.3 \\
$\nu$               & 0.27   &  0.27   &  0.27  &  &  0.34   &  0.34  &   0.34  & &  0.33  & 0.34  & 0.34 \\
$v_{\text{L}}$      &  7128  &  7168   &  7140  &  &  6368   &  6393  &   6396  & &  6387  & 6411  & 6419  \\ 
$v_{\text{T}}$      &  4014  &  4043   &  4012  &  &  3148   &  3166  &   3165  & &  3187  & 3180  & 3194  \\
$\bar{v}$           &  4466  &  4497   &  4465  &  &  3534   &  3554  &   3553  & &  3575  & 3569  & 3584  \\
$\theta_{\text{D}}$ & 595.9  &  601.8  &  595.3 &  &  470.3  &  474.3 &   472.4 & &  475.7 & 476.3 & 476.5 \\
$H_{\text{V}} $     & 13.65  &  13.96  &  13.50 &  &  5.99   & 6.11   &   6.05  & &  6.29  & 6.19  & 6.24  \\
$B/G$               & 1.82   &  1.81   &  1.83  &  &  2.76   &  2.75  &   2.75  & &  2.68  & 2.73  & 2.71 \\
$A_{B}$             &   0.0  &   0.0   &  0.0   &  &  0.45   &  0.45  &  0.48   & &  0.51  &  0.54 & 0.45  \\
$A_{G}$             &   1.23 &  1.37   &  1.19  &  &  7.23   &  7.36  &  6.82   & &  6.24  &  6.74 &  6.30  \\
$A^{\text{U}}$      &  0.12  &  0.14   &  0.12  &  &  0.79   &  0.80  &  0.74   & &  0.68  &  0.73 & 0.68    \\

\hline \hline
\end{tabularx}
\end{table}  

The knowledge of single-crystal elastic constants $C_{ij}$ allows one to evaluate mechanical and other important properties of polycrystals. The list of the parameters which describe the polycrystalline behaviour of LaNiO$_3$ is given in Table \ref{tab3}. For the sake of convenience, Table \ref{tab3} presents values of bulk and shear moduli obtained by applying not only the averaging method recommended by Hill ($B$, $G$), but also the initial isostrain and isostress approximations correspondingly suggested by Voigt ($B_{\text{V}}$, $G_{\text{V}}$) and Reuss ($B_{\text{R}}$, $G_{\text{R}}$). It is worth to note a very good agreement between the bulk modulus $B$ and the one evaluated from the BM equation of state $B_{\text{BM}}$ indicating a high numerical accuracy of our calculations. Regarding the available experimental data, a value of $B^{\text{Rhomb.}}_{\text{Expt.}}\approx188$ GPa from the room-temperature measurements [\onlinecite{zhou_52}] nicely corresponds to $B^{\text{Rhomb.}}\approx200$ GPa, especially if one would take into consideration thermal effects which tend to reduce the elastic moduli. A comparison between the different phases of LaNiO$_3$ in Table \ref{tab3} reveals that all three structures exhibit a similar level of compressibility because the difference in their bulk moduli $B$ does not exceed $\sim8\%$. However, the cubic structure is more resistive to shear and uniaxial deformations: $G^{\text{Cub.}}\approx119$ GPa against $G^{\text{Rhomb.}}\approx73$ GPa and $G^{\text{Mon.}}\approx74$ GPa, and $E^{\text{Cub.}}\approx301$ GPa against $E^{\text{Rhomb.}}\approx194$ GPa and $E^{\text{Mon.}}\approx197$ GPa. This fact is also reflected in the lower value of Poisson's ratio and higher values of sound velocities and Debye temperature compared to the rhombohedral and monoclinic phases. The available experimental data for Debye temperature $\theta_{\text{D}}^{\text{Expt.}}=420$ K [\onlinecite{rajeev_53}] indicate a fairly good agreement with calculations $\theta_{\text{D}}^{\text{Rhomb.}}\approx472$ K, since the deviation from experiment does not exceed $\sim13\%$. Further analysis of elastic parameters shows that the Vickers hardness for cubic LaNiO$_3$ is more than two times larger in comparison to the rhombohedral and monoclinic LaNiO$_3$: $H_{\text{V}}^{\text{Cub.}}\approx13.7$ GPa against $H_{\text{V}}^{\text{Rhomb.}}\approx6.1$ GPa and $H_{\text{V}}^{\text{Mon.}}\approx6.2$ GPa, respectively. This well-known mechanical property, defined as the ability of a material to resist against being dented or scratched by another, was obtained according to the expression $H_{\text{V}}=0.92(G/B)^{1.137}G^{0.708}$ recently proposed by Tian $et$ $al$. [\onlinecite{tian_50}]. Interestingly, the apparent distinction in hardness does not change the ductile nature of cubic, rhombohedral, and monoclinic LaNiO$_3$, although the latter two structures demonstrate a more pronounced ductility with corresponding $B/G$ ratios of $\sim2.75$ and $\sim2.71$ compared to $\sim1.82$ of the cubic phase. The most common critical value that separates ductile and brittle character was introduced by Pugh [\onlinecite{pugh_51}]: if $B/G > 1.75$, the material behaves in a ductile manner, otherwise, its behaviour should be related with brittleness. 

Elastic anisotropy can have considerable effects on durability of the material through the formation of microcracks, therefore this parameter plays an important practical role not less than the others. The percentage anisotropy in compressibility and shear for polycrystals, originally suggested by Chung and Buessem [\onlinecite{chung_54}], is defined as $A_{B}=100(B_{\text{V}}-B_{\text{R}})/(B_{\text{V}}+B_{\text{R}})$ and $A_{G}=100(G_{\text{V}}-G_{\text{R}})/(G_{\text{V}}+G_{\text{R}})$, respectively. Here, a value of zero represents elastic isotropy, while a value of $100\%$ identifies the largest possible anisotropy. From Table \ref{tab3} it is seen that rhombohedral and monoclinic phases of LaNiO$_3$ exhibit a similarly weak anisotropy in compressibility with $A_{B}^{\text{Rhomb.}}\approx0.46\%$ and $A_{B}^{\text{Mon.}}\approx0.5\%$, but their anisotropy in shear appears to be more pronounced, as the numbers accordingly increase up to $A_{G}^{\text{Rhomb.}}\approx7.14\%$ and $A_{G}^{\text{Mon.}}\approx6.43\%$. In the meantime, the perfect isotropy in compressibility for the cubic phase accompanied by its relatively weak anisotropy in shear $A_{G}^{\text{Cub.}}\approx1.26\%$ speaks for the highest degree of elastic isotropy among the considered structures of LaNiO$_3$. And indeed, the following finding is confirmed by the universal anisotropy index $A^{\text{U}}=5(G_{\text{V}}/G_{\text{R}})-(B_{\text{V}}/B_{\text{R}})-6$ [\onlinecite{ranganathan_55}], the values of which can also be found in Table \ref{tab3}. By taking into account both the bulk and the shear contributions, $A^{\text{U}}$ represents a universal measure to quantify the single-crystal elastic anisotropy. For an isotropic crystal, $A^{\text{U}}=0$ and deviations from zero determine the extent of elastic anisotropy. Gladly, the comparison of calculated values $A^{\text{U}}_{\text{Cub.}}\approx0.13$, $A^{\text{U}}_{\text{Rhomb.}}\approx0.78$, and $A^{\text{U}}_{\text{Mon.}}\approx0.7$ shows a full compatibility with our previous observations drawn from $A_{B}$ and $A_{G}$.    

In general, one can notice that polycrystalline parameters derived for rhombohedral and monoclinic phases of LaNiO$_3$ bear a very close resemblance to each other, since in most cases the discrepancy does not exceed $\sim10\%$. On one hand, it implies that transformation of rhombohedral LaNiO$_3$ into monoclinically distorted structure induced by compressive or tensile strains does not have a strong influence on its mechanical behaviour. But on the other hand, one should take into consideration that here we are dealing with a fully relaxed monoclinic system which is slightly different from non-relaxed ones experimentally identified on SrTiO$_3$ and LaAlO$_3$ substrates [\onlinecite{may_18}]. However, we believe that our elastic parameters pretty reliably describe monoclinic LaNiO$_3$ as long as the thickness of the film is sufficiently large to prevent from the occurrence of metal-to-insulator transition. Experimental investigation of electrical resistivity [\onlinecite{son_56}] and electronic structure [\onlinecite{gray_57}] has revealed that films that are thick enough (10 nm or somewhat less depending on the substrate) show clearly pronounced metallic behaviour that is typical for bulk LaNiO$_3$ and can be properly represented at DFT level. But for the ultrathin LaNiO$_3$ films composed of only several monolayers, a metal-to-insulator transition is observed and a standard DFT scheme fails to reproduce it [\onlinecite{gray_57}]. Thus, in order to accurately describe the elastic properties of ultrathin LaNiO$_3$ structures, one should apply more sophisticated theoretical methods. In other cases, our given data seem to be sufficient.   

\section{Conclusions}

Elastic properties allow one to evaluate many important mechanical and thermodynamic features of the material, thus in this paper we remedy the existing gap of knowledge by presenting single-crystal elastic constants and deduced polycrystalline parameters for the metallic perovskite LaNiO$_3$. For the sake of greater accuracy, a bunch of DFT approaches has been tested to reproduce the experimentally observed low-temperature crystalline and electronic structures of bulk LaNiO$_3$. As the revised functionals for solids have demonstrated the best performance, they were chosen for the next calculations of elastic properties for the most common rhombohedral as well as high-temperature cubic and strain-induced monoclinic phases of LaNiO$_3$. The obtained results indicate that all three structures are mechanically stable, behave in a ductile manner, and exhibit a similar level of compressibility. However, rhombohedral and monoclinic LaNiO$_3$ are softer against uniaxial and shear deformations in comparison to the cubic phase, and this fact is somewhat reflected in the higher values of the Poisson's ratio and lower ones for the Vickers hardness, sound velocities, and Debye temperature. Concerning the elastic isotropy, it can be stated that, in general, LaNiO$_3$ is more anisotropic in shear than in compression but the overall anisotropy is not strongly pronounced. It is also worth to note that rhombohedral and monoclinic phases demonstrate a very close polycrystalline behaviour which in turn implies that LaNiO$_3$ retains its mechanical properties when transforming from thin film to bulk material. We expect the obtained data and proposed insights to be useful for various practical applications and possible future experiments.     

\begin{acknowledgments}
The authors are thankful for the HPC resources provided by the ITOAC of Vilnius University. 
\end{acknowledgments}

\providecommand{\noopsort}[1]{}\providecommand{\singleletter}[1]{#1}%

\end{document}